\documentclass[12pt]{article}
\usepackage{epsf}
\font\amsmathe=msbm10 scaled \magstep1

\def\SL{\sum\limits}
\def\nn{\nonumber\\}
\def\tr{{\rm tr}}
\def\1{\mbox{\rm 1\kern-.25eml}}

\def\s{\mbox{\boldmath$s$}}
\def\e{\mbox{\boldmath$e$}}
\def\endproof{\hfill\rule{2mm}{2mm}}
\def\P{\mbox{\boldmath$\pi$}}
\def\x{\mbox{\bf x}}
\def\la{\left\langle}
\def\ra{\right\rangle}

\newcommand{\bee}{\begin{equation}}
\newcommand{\ee}{\end{equation}}
\newcommand{\bea}{\begin{eqnarray}}
\newcommand{\eea}{\end{eqnarray}}
\def\R{\hbox{\amsmathe R}}

\newcommand{\N}{{\rm I\kern-.16em N}}
\def\Z{\hbox{\amsmathe Z}}
\newcommand{\NP}{{\sl Nucl. Phys.}}

\newcommand{\PRL}{{\sl Phys. Rev. Lett.}}

\bigskip 
\nopagebreak
\begin{document}
%%%%%%%%%%%%%%%%%%%%  Version Sep 30, 2002 %%%%%%%%%%%%%%%%%%%%%%%%%
\begin{flushright}
MPI-PhT/02-50\\
\end{flushright}
\bigskip\bigskip
\begin{center}
{\Large Critical behavior in a quasi D dimensional spin model
\footnote{This work is based to a large extent on the second named 
author's 
doctoral dissertation\cite{Y}}
}
\end{center}
\centerline{Erhard Seiler and Karim Yildirim}
\vskip5mm
\centerline{\it Max-Planck-Institut f\"ur Physik}
\centerline{\it (Werner-Heisenberg-Institut)}
\centerline{\it F\"ohringer Ring 6, 80805 Munich, Germany}
\centerline{\it e-mail:ehs@mppmu.mpg.de}
\bigskip \nopagebreak
\begin{center}
{\it Dedicated to the memory of Adrian Patrascioiu}
\end{center}

\begin{abstract}
We study a classical spin model (more precisely a class of models)
with $O(N)$ symmetry that can be viewed as a simplified $D$ dimensional 
lattice model. It is equivalent to a non-translationinvariant one 
dimensional model and contains the dimensionality $D$ as a parameter that 
need not be an integer. The critical dimension turns out to be 2, just as 
in the usual translation invariant models. We study the phase structure, 
critical phenomena and spontaneous symmetry breaking. Furthermore we 
compute the perturbation expansion to low order with various boundary 
conditions.
In our simplified models a number of questions can be answered
that remain controversial in the translation invariant models, such
as the asymptoticity of the perturbation expansion and the role of
super-instantons. We find that perturbation theory produces the right 
asymptotic expansion in dimension $D\leq 2$ only with special boundary 
conditions. Finally the model allows a test of the percolation ideas of 
Patrascioiu and Seiler.
\end{abstract}
\newpage
%\pacs{11.25.Bt, 11.15.Ha, 75.10.Hk}
    
%%%%%%%%%%%%%%%%%%%%  Version March 11, 1999 %%%%%%%%%%%%%%%%%%%%%%%%%
\vskip4mm \noindent
\section{Introduction}
\setcounter{equation}{0}

In their well-known paper on the Mermin-Wagner theorem \cite{DS},
Dobrushin and Shlosman considered in a side remark a model that is the
prototype of the model studied in this paper. Their purpose was to 
illustrate the importance of the smoothness of the interaction for the 
question of symmetry breaking. Here we take advantage of the fact that 
this model on the one hand has almost the simplicity of a one-dimensional
model but on the other hand has a tunable parameter $D$ playing the
role of dimension. For integer values of $D$, the model can actually
be implemented as a model on the lattice $\Z^D$ with nontranslational
Hamiltonian.

Such a model was also studied in \cite{PSperc1} in order to verify that 
the smoothness restrictions of \cite{DS} can be relaxed and are only 
needed in a neighborhood of the minima of the Hamiltonion. Finally, in 
Georgii's book \cite{georgii} a similar non-translation invariant chain,
but with Ising spins is analyzed.

Here we use this type of model as a laboratory to test various ideas
proposed by Patrascioiu and the first named author in their quest to
prove that the conventional distinction between abelian and nonabelian
is unjustified. The plan of the paper is as follows: After giving the 
definitions of the model and the various boundary conditions used, in 
Section 3 we use the full, nonperturbative solution of the model to 
study the phase structure as a function of the `dimension' $D$. It
turns out that the critical dimension is still $D=2$; for $D\leq 2$ the 
model does not show spontaneous magnetization or phase coexistence, 
whereas for $D>2$ is does. For $D=2$ the model (without external magnetic 
field) and does not show asymptotic freedom.

In Section 4 we contrast those nonperturbative results with the results
of (low order) perturbation theory (PT). We find that for zero magnetic
field PT at the level of one loop already becomes dependent on the 
boundary conditions (b.c.) used for all $D\leq 2$ and therefore in general 
does not yield the correct asymptotic expansion of the model. The 
analogous result for the $1D$ model is well known (\cite{BR,Pprl54}). In 
this simple class of models it is, however, easy to find b.c. in which 
local observables are independent of the size $L$ of the system and hence 
PT of the finite system does give the correct asymptotic expansion. Such 
b.c. exist in principle also for the full translation invariant models. 
Formally they arise by integrating out all the variables outside a box of 
suitable size, more precisely these b.c. are obtained using the 
Dobrushin-Lanford-Ruelle (DLR) equations \cite{DLR,BS}. In our casefor 
$1\leq D\leq 2$ those DLR boundary conditions turns out to be simply the 
standard free b.c.. Whereas the nonperturbative results are rather easily 
obtained for our model, PT turns out to be harder to compute; for this 
reason we limit ourselves to one loop. This is sufficient, however, 
for seeing all the effects and subtleties we are interested in.

Finally, in Section 5, we discuss some percolation properties of our
model. In particular we test the ideas of \cite{PSperc2,PSperc3} on the
percolation properties of various sets defined by spins pointing in
certain subsets of the spheres $S_{N-1}$. We find that in $1\leq D\leq 2$
we are generally in the situation of `critical percolation', as suggested
by \cite{PSperc2,PSperc3}.

We should stress that our analysis shows that there is no qualitative
difference between the abelian case $N=2$ and the nonabelian one ($N>2$).
In that sense it lends support to the `heretical' scenario of Patrascioiu
and Seiler that predicts the existence of a `soft' phase in all $2D$
$O(N)$ models. Sceptics might still argue, however, that the model is more
ordered than the standard translation invariant ones and that this is the
reason for the existence of the soft phase.

In this paper we are to a large extent presenting results of \cite{Y}, 
with emphasis on the physical interpretation. For more mathematical and
computational details we refer the reader to that work.

\section{The Model}
\setcounter{equation}{0}
\subsection{Definition}

To motivate the model, we start with a general class of classical $O(N)$ 
spin models defined on lattices $\Z^D$, but with link dependent 
couplings. To each lattice site $x\in \Z^D$ we associate a classical 
$O(N)$ spin, i.e.
\bee
x \to \s(x), \ \ \s(x)\in \R^N, \ \ ||\s(x)||=1
\ee

The Hamiltonian is of the form
\bee
H=-\sum_{\langle xy\rangle} \beta_{xy} \s(x)\cdot \s(y)-h\sum_x s_N(x)
\ee
where $s_N(j)=\s(j)\cdot\e_N$ and $\e_N$ is the unit vector in $\R^N$ 
pointing in the $N$th direction; the sum is over nearest neighbors and 
$h$ represents an external magnetic field; the dot denotes the standard 
euclidean scalar product in $\R^N$.

\begin{figure}[ht]
\begin{center}
\unitlength2.0cm
\begin{picture}(3,3)
\multiput(0,0)(0,0.5){7}{\circle*{0.2}}
\multiput(0.5,0)(0,0.5){7}{\circle*{0.2}}
\multiput(1,0)(0,0.5){7}{\circle*{0.2}}
\multiput(1.5,0)(0,0.5){7}{\circle*{0.2}}
\multiput(2,0)(0,0.5){7}{\circle*{0.2}}
\multiput(2.5,0)(0,0.5){7}{\circle*{0.2}}
\multiput(3,0)(0,0.5){7}{\circle*{0.2}}
\multiput(0,0)(0.5,0){7}{\line(0,1){3}}
\multiput(0,0)(0,0.5){7}{\line(1,0){3}}
%\thicklines
\linethickness{0.07cm}
\multiput(0,0)(3,0){2}{\line(0,1){3}}
\multiput(0,0)(0,3){2}{\line(1,0){3}}
\multiput(0.5,0.5)(2,0){2}{\line(0,1){2}}
\multiput(0.5,0.5)(0,2){2}{\line(1,0){2}}
\multiput(1,1)(1,0){2}{\line(0,1){1}}
\multiput(1,1)(0,1){2}{\line(1,0){1}}
\thinlines   
%\put(1.55,1.6){$\tilde{g}_1$}
%\put(0.05,1.7){$\tilde{g}_4$}
%\put(0.7,2.05){$\tilde{b}_2$}
%\put(2.7,0.55){$\tilde{b}_3$}
\end{picture}
\end{center}
\caption{Scheme of the lattice for $D=2$; on the thick lines we send
$\beta_{xy}\to\infty$}
\label{lattice}
\end{figure}
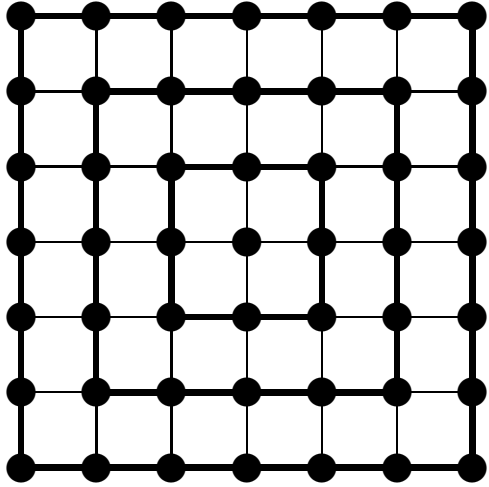

We now choose an origin and surround it with a family of concentric 
quadratic (cubic, hypercubic) `shells' (see Fig.\ref{lattice}). We 
`freeze' all the links (nearest neighbor pairs) sitting inside one of the
shells by sending the corresponding $\beta_{xy}\to\infty$, thereby 
forcing all spins within such a shell to be equal. All other couplings 
$\beta_{xy}$ are set equal to a constant $\beta$.

We can therefore identify all the points within a `shell'; thus the 
resulting model can be equivalently described as a non
translation invariant semi-infinite spin chain formally defined by
the Hamiltonian
\bee
H=-\sum_{j=1}^\infty (b_j \s(j)\cdot \s(j+1)-h_j s_N(j))
\ee
where
\bee
b_j=2D(2j-1)^{D-1}, 
\ \ j=1,2,3,...
\ee
\bee
 h_j=h g_j
\ee
with
\bee
g_j=\lbrack (2j-1)^D-(2j-3)^D\rbrack,
\ \ j=2,3,4,...
\ee
and 
\bee
g_1=1.
\ee
Note that for $j\to\infty$ both $b_j$ and  $g_j$ behave asymptotically		
like $2D(2j)^{D-1}$.

As usual, we will have to study first finite chains of length $L$ and
then take the thermodynamic limit $L\to\infty$. This will require imposing
certain b.c. at the end of the chain. In addition we will also generalize
the model by introducing similar b.c. at the beginning of the chain. The
form of the Hamiltonian given above corresponds to free b.c. at the
beginning of the chain, which is natural from the point of view of the
$D$-dimensional lattice, but we will also be interested in posing
Dirichlet b.c. at the origin, which in combination with Dirichlet b.c. at
$j=L$ will correspond to the superinstanton b.c. introduced in
\cite{PSsi}. A class of b.c. that allows to interpolate between free b.c.
and Dirichlet b.c. is given by the following
finite volume Gibbs measures:
\bee
d\mu_L={1\over Z_L} {\rm e}^{-\beta H_L} \prod_{j=1}^Ld\nu(\s(j))
\ee
where $d\nu$ is the $O(N)$ invariant probablity measure on $S_{N-1}$ and
\bee
H_L=-\sum_{j=1}^{L-1} b_j \s(j)\cdot \s(j+1)-\sum_{j=1}^L h_j s_N(j))
-as_N(1))-bs_N(L))
\label{bc}
\ee
Free b.c. at $j=1$ ($j=L$) are given by putting $a=0$ ($b=0$);
Dirichlet b.c. at $j=1$ ($j=L$) are obtained by taking the limit
$a\to\infty$ ($b\to\infty$). Putting $b=b_L$, however, also corresponds
to Dirichlet b.c., but at $j=L+1$ for a chain of length $L+1$.
Likewise $a=1$ can be interpreted as Dirichlet b.c. on an extended chain.

We will also make use of more general b.c. at $j=L$ which are obtained
by replacing the factor $\exp(\beta bs_N(L))d\nu(\s(L))$ in the
Gibbs measure for a finite chain with a general positive measure written
formally (in distributional notation) as $\tilde\psi(\s(L))d\nu(\s(L))$.
We may even let $\tilde\psi$ depend on the length $L$ of the chain; in
particular this is necessary if we want to describe Dirichlet b.c. at
$j=L+1$ by choosing $b=b_L$, as described above.

The model we have defined contains the dimension $D$ as a parameter
which no longer has to be an integer (it could even be chosen complex).
In the following we will treat $D$ as a real parameter $\geq 1$ and we 
will use almost exclusively the representation of the model on a chain.

Let us note one crucial fact which turns out to be responsible for the 
dependence of the properties of the model on the `quasidimension' $D$:

{\bf Proposition 2.1:} For $1\leq D\leq 2$ and any $k>0$ \ \
$\sum_{j=k}^L b_j^{-1}\to\infty$ for $L\to\infty$, whereas for 
$D>2$\ \ $\sum_{j=k}^L b_j^{-1}$ converges to a finite limit. More
precisely, for $D<2$\ \  $\sum_{j=k}^L b_j^{-1}=O(k^{2-D})$, for 
$D=2$ it is $O(\ln k)$, and for $D>2$ it approaches its limit like 
$k^{2-D}$.

The proof is a straightforward consequence of the asymptotics of the
$b_j$.

\subsection{Thermodynamic quantities}

To define thermodynamic quantities like free energy density, 
magnetization or susceptibility, we first note that the volume of the
chain of length $L$ is naturally defined as
\bee
V_L=(2L+1)^D.
\ee
The free energy density then becomes
\bee
f_L=-{1\over \beta V_L}\ln Z_L
\ee
and the magnetization we define as usual as
\bee
M_L=-{\partial f_L\over\partial h}
\ee
Thus
\bee
M_L={1\over V_L}\sum_{j=1}^Lg_j\la s_N(j)\ra .
\label{magn}
\ee
Likewise we define the (longitudinal) susceptibility as
\bee
\chi_L={\partial M_L\over \partial h}=-{\partial^2f\over \partial h^2}
\ee
which means
\bee
\chi_L={\beta\over V_L}\sum_{j,l=1}^L g_jg_l\biggl[\la s_N(j)s_N(l)\ra
-\la s_N(j)\ra\la s_N(l)\ra\biggr]
\label{susc}
\ee
All these definitions are analogous to those in translation
invariant systems. Of course we are mostly interested in the 
thermodynamic limit $L\to\infty$. We use the definitions (\ref{magn}) and
(\ref{susc}) to define the spontaneous magnetization and the
susceptibility also for the thermodynamic limit at $h=0$.

\subsection{Superinstantons}

For $D\leq 2$ the model has superinstantons, just like the translation
invariant model \cite{PSsi}. These are configurations of arbitrarily
low energy which are disordering the system and are responsible for the 
absence of spontaneous symmetry breaking (Mermin-Wagner theorem \cite{MW}).

More concretely, these configurations are obtained by imposing b.c. at
$j=i$ and $j=k$ such that $\s(i)\cdot\s(k)=\arccos(\alpha)<1$ and
minimizing the energy under those b.c.. The existence of such a
minimizing configuration is obvious, because the energy is a
continuous function on a compact set. It is clear that in this minimal 
configuration the spin will vary on a great circle in $S_{N-1}$, so we 
can describe the configuration by the angle 
\bee
\phi_j=\arccos\left(\s(j)\cdot\s(j+1)\right).
\ee
The quantitiy to be minimized is then
\bee
E(i,k)=-\sum_{j=i}^{k-1}b_j(\cos(\phi_j)-1)
\ee
under the condition that
\bee
\sum_{j=i}^{k-1}\phi_j=\alpha
\ee
Using a Lagrange multiplicator $\lambda$ one obtains therefore the
equations
\bee
\lambda-b_j\sin(\phi_j)=0
\label{lagr}
\ee
which has the solutions
\bee
\phi_j=\arcsin({\lambda\over b_j})
\ee
and yields for the minimizing configuration
\bee
E_{s.i.}(i,k)=\sum_{j=i}^{k-1} 
b_j\Biggl(1-\sqrt{1-{\lambda^2\over b_j^2}}\Biggr)
\ee
with
\bee
\sum_{j=i}^{k-1}\arcsin\bigl({\lambda\over b_j}\bigr)=\alpha .
\ee
We will now derive upper and lower bounds for $E_{s.i.}(i,k)$. Using
\bee
1-x\leq\sqrt{1-x}\leq 1-{x\over 2}\ \ {\rm for} \ \ 0\leq x\leq 1
\ee
we obtain
\bee
{1\over 2}\sum_{j=i}^{k-1}{\lambda^2\over b_j}\leq E_{s.i.}(i,k)
\leq \sum_{j=i}^{k-1}{\lambda^2\over b_j}
\label{en}
\ee
Bounds for the Lagrange multiplier $\lambda$ are obtained using
\bee
x\leq\arcsin(x)\leq 2x \ \ {\rm for} \ \ 0\leq x\leq 1
\ee
from which we obtain, using eq.(\ref{lagr})
\bee
\lambda \sum_{j=i}^{k-1} b_j^{-1}\leq \alpha \leq
2 \lambda \sum_{j=i}^{k-1}b_j^{-1}
\ee
or
\bee
{\alpha\over 2 \sum_{j=i}^{k-1}b_j^{-1}}\leq\lambda \leq
{\alpha\over \sum_{j=i}^{k-1}b_j^{-1}} .
\ee
Combining this with eq.(\ref{en}) we finally obtain
\bee
{\alpha^2\over 8 \sum_{j=i}^{k-1}b_j^{-1}}\leq E_{s.i.}(i,k)\leq
{\alpha^2\over \sum_{j=i}^{k-1}b_j^{-1}}
\ee
Here we can see again the distinction between low ($D\leq 2$) and high
$(D>2$) quasidimension:\\
For $D\leq 2$
\bee
\lim_{k\to\infty} E_{s.i.}(i,k)=0
\ee
whereas for $D>2$ $E_{s.i.}(i,k)$ is uniformly in $k$ bounded away from 0:
\bee
E_{s.i.}(i,k)\geq {\alpha^2\over 8 \sum_{j=i}^\infty b_j^{-1}}>0.
\label{lb}
\ee
The fact that the superinstantons in $D\leq 2$ cost arbitrarily little
energy is responsible for the fact that the system has no long range
order (just as the translation invariant models are according to the
Mermin-Wagner theorem \cite{MW}). This will be proven in the next section.

Due to the freezing of all the spins within one `layer', the model does
not have instanton-like configurations.
    %The Model
\section{Nonperturbative Analysis}
\setcounter{equation}{0}

\subsection{Transfer Matrices}

Because our model only couples nearest neighbors, it can be described
easily in terms of transfer operators, which are of course site
dependent. These transfer `matrices' are trace class operators on the 
Hilbert space 
\bee{\cal H}=L^2(S_{N-1},d\nu),\ee 
which we normalize such that the largest eigenvalue equals $1$. In slight 
abuse of notation, we will use the same symbol for the operators and 
their integral kernels.

The normalized transfer matrix from site $j$ to site $j+1$ is
\bee
\tilde{\cal T}_j(\s(j),\s(j+1))= {1\over z_j} \exp\biggl[
\beta \bigl(b_j\s(j)\cdot\s(j+1)
+h_js_N(j)\bigr)\biggr]
\ee
with $z_j$ chosen such that
\bee
||\tilde{\cal T}_j ||=1
\ee
These transfer operators are not self-adjoint, but it is easy to transform
them by a similarity transformation with a bounded operator into the 
self-adjoint operators with integral kernel
\bee
{\cal T}_j(\s(j),\s(j+1))= {1\over z_j} \exp\biggl[
\beta \bigl(b_j\s(j)\cdot\s(j+1)
+{h_j\over 2}(s_N(j)+s_N(j+1)\bigr)\biggr]
\ee
These operators are in fact positive, as can be seen easily be expanding 
$\exp\bigl[\s(j)\cdot\s(j+1)\bigr]$.

Since the transfer operators have also positive integral kernels, by a 
trivial change of normalization it is possible to interprete them as
transition probabilities and the whole system as a Markov chain. 
This point of view will, however, not play a great role in this paper.

We introduce a shorthand for products of transfer matrices:
\bee
{\cal T}_{jk}\equiv \prod_{j=1}^{k-1}{\cal T}_j
\ee
and define
\bee
\beta_j\equiv\beta b_j
\ee

Using the transfer matrices, it is possible to rewrite for instance
the expectation value of a spin component at site $k$
\bee
\la s_a(k)\ra_L\equiv\int d\mu_L s_a(i)
\ee
in the form
\bee
\la s_a(i)\ra_L={(\psi, {\cal T}_{1i} s_a{\cal T}_{iL} \tilde\psi)
\over (\psi, {\cal T}_{1L} \tilde\psi)}
\label{1ptfct}
\ee
where we denote by $(.,.)$ the scalar product in ${\cal H}$. $\psi$ and
$\tilde\psi$ are suitable vectors in ${\cal H}=L^2(S_{N-1})$ describing
the b.c. (actually we may even replace them by general positive measures 
on $S_{N-1}$). The spin variable $s_a$ appearing on the right hand side
of eq.(\ref{1ptfct}) is to be interpreted as a multiplication operator 
on $L^2(S_{N-1})$.

The spin-spin two point function 
\bee
\la s_a(i)s_b(k)\ra_L\equiv\int d\mu_Ls_a(i)s_b(k)
\label{2ptfct}
\ee
for $i\leq k$ can similarly be expressed as
\bee
\la s_a(i)s_b(k)\ra_L={(\psi, {\cal T}_{1i}
\s_a{\cal T}_{ik}\s_b{\cal T}_{kL}\tilde\psi)
 \over (\psi, {\cal T}_{1L}\tilde\psi)}
\ee

For a nonvanishing magnetic field we do not expect any interesting 
phenomena; the transfer matrices will force the spins `at infinity' to
be aligned with the direction of the magnetic field $\e_N$, therefore
one expects a unique thermodynamic limit, independent of the b.c. imposed 
at $L$.

For vanishing magnetic field the situation is more interesting; therefore
from now on we will assume $h=0$. To analyze the situation, we need some 
preparation (see for instance \cite{Cu,SY}):

The Hilbert space ${\cal H}=L^2(S_{N-1},d\nu)$ can be decomposed into
the eigenspaces ${\cal H}_l$ of the Laplace-Beltrami operator
$\Delta_{LB}$ on the sphere $S_{N-1}$:
\bee
{\cal H}= \bigoplus_{l=0}^\infty {\cal H}_l
\ee
The projections ${\cal P}_l$ onto the eigenspaces are integral operators
which for $N>2$ can be expressed in terms of the Gegenbauer polynomials 
$C_l$ (see \cite{mag,vilenkin})
\bea
C_l^{\frac{N}{2}-1}(x)&=&\frac{1}{\Gamma\left(\frac{N}{2}-1\right)}
\SL_{m=0}^{\left[\frac{l}{2}\right]}(-1)^m\frac{\Gamma\left(\frac{N}{2}-1+l-m
\right)}{m!\,(l-2m)!}(2x)^{l-2m},\quad N>2,\nn
C_l^0(x)&=&\SL_{m=0}^{\left[\frac{l}{2}\right]}(-1)^m\frac{\Gamma(l-m)}{
\Gamma(m+1)\Gamma(l-2m+1)}(2x)^{l-2m},\quad l\neq 0.
\eea
as follows:
\bee
{\cal P}_l(\s, \s^\prime)={2l+N-2\over N-2} C_l^{{N\over 2}-1}
(s\cdot s^\prime)
\ee
For $N=2$ the Gegenbauer polynomials have to be replaced by the Chebyshev
polynomials of the first kind:
\bee
T_o(x)=1
\ee
\bee
T_l(x)=\lim_{N\to 2}{1\over N-2} C_l^{{N\over 2}-1}(x) \ \  {\rm for} l>1.
\ee 
In the following we will write the equations in the form valid for $N>2$, 
involving the Gegenbauer polynomials, with the understanding that for 
$N=2$ the analogous formulae involving the Chebyshev polynomials hold.

Due to the $O(N)$ invariance of the transfer matrices for $h=0$, all
${\cal T}_j$ commute with those projections, and the spaces ${\cal H}_l$
are also simultaneous eigenspaces of all ${\cal T}_j$. Because the
Gegenbauer (Chebyshev) polynomials form a complete orthogonal set on the 
interval $[-1,1]$, the integral kernel of the transfer matrix can be 
expanded in the sense of $L^2([-1,1],(1-x^2)^{N-2\over 2} dx)$ as
\bee
{\cal T}_j(\s,\s^\prime)
=\sum_{l=0}^\infty c_l C_l^{N/2-1}(\s\cdot\s^\prime)
\ee
From this we see that each subspace ${\cal H}_l$ is an eigenspace with
eigenvalue $\lambda_l$ of ${\cal T}_j$. The eigenvalues 
$\lambda_l$ are therefore given by
\bea
\lambda_l(\beta_j)&=&{\tr {\cal T}_j {\cal P}_l\over \tr{\cal 
P}_l}=
{1\over z_j}\int_{-1}^1 dx \exp(\beta_jx)\bigl(1-x^2\bigr)^{{N-3\over 2}}
{C_l^{n/2-1}(x)\over C_l^{n/2-1}(1)}\cr
&=&{1\over z_j}\sqrt{\pi} \Gamma\bigl({N-1\over 2}\bigr) 
\biggl({2\over \beta_j}\biggr)^{N-2\over 2} I_{l+{N\over 2}-1}(\beta_j)
\eea
where $I_r(.)$ is the modified Bessel function (\cite{mag}).
From the fact that the ${\cal T}_j$ are positivity improving operators
(see \cite{RS}) it follows that
\bee
0<\lambda_l(\beta_j)<\lambda_o(\beta_j)=1.
\ee
The normalization $\lambda_o(\beta_j)=1$ allows to determine $z_j$ 
explicitly and we get for the eigenvalues finally
\bee
\lambda_l(\beta_j)={I_{l+{N\over 2}-1}(\beta_j)\over I_{{N\over 
2}-1}(\beta_j)}
\ee

\subsection{Thermodynamic limit}

In this subsection we will discuss explicitly the thermodynamic limit
for the one and two point functions of the spins. The generalization
to higher $n$-point functions is straightforward in principle. 

To take the thermodynamic limit we need some information about the limit 
of the product of a large number of transfer matrices. Since for $h=0$ 
all transfer matrices commute and have the same eigenprojections 
${\cal P}_l$, we obtain
\bee
{\cal T}_{ik}=\sum_{l=0}^\infty\prod_{j=i}^{k-1}\lambda_l(\beta_j)
{\cal P}_l .
\ee
So we have to study the behavior of the products
$\prod_{j=i}^{k-1}\lambda_l(\beta_j)$ for large $k$. For this purpose we 
can use some results of \cite{SY, Cu}, where we analyzed the asymptotic 
behavior of such eigenvalues (actually for more general forms of the 
Hamiltonian). The main result is
\bee
\ln\lambda_l(\beta_j)=1-{l(l+N-2)\over 2 \beta_j}+O(\beta_j^{-2})
\label{asymp}
\ee
We define for $i\le k$\\
\bee
\Lambda_l(i,k)=\prod_{j=i}^{k-1} \lambda_l(\beta_j)
\ee
Using (\ref{asymp}) and Prop. 2.1 we find the asymptotic behavior 
of $\ln\Lambda_l(1,k)$:
\bee
\ln\Lambda_l(1,k)=-{l(l+N-2)\over 2\beta}
{1\over 2^D D(2-D)} k^{2-D}+O\left({k^{2(2-D)}\over\beta^2}\right) 
\ \ {\rm for}\ \  1\leq D<2,
\ee
\bee
\ln\Lambda_l(1,k)=-{l(l+N-2)\over 2\beta}{1\over 8}{\ln k}
+O(\beta^{-2}) \ \ {\rm for}\ \ D=2
\ee
and
\bee
\ln\Lambda_l(k,\infty)=-{l(l+N-2)\over 
2\beta}{1\over 2^D D(D-2)} k^{2-D}+O\left({k^{2(2-D)}\over\beta^2}\right)
 \ \ {\rm for}\ \ D>2.
\label{asympt}
\ee
Using this, we obtain easily

\noindent
{\bf Proposition 3.1:} For $D\leq 2$ ${\cal T}_{i\infty}\equiv
s$-$\lim_{k\to\infty}{\cal T}_{ik} ={\cal P}_0$. For $D>2$
${\cal T}_{i\infty}$ exists but is strictly positive;
$s$-$\lim_{i\to\infty} {\cal T}_{i\infty}=\1$. 

{\it Proof:}
To prove strong convergence, it suffices by a $3\epsilon$ argument
to prove convergence of all the eigenvalues. We now treat three cases
separately:\\
(1) $D=1$:
All ${\cal T}_j$ are equal and the result follows
trivally from the fact that for $l\geq 1$ $\lambda_l<1$
\\
(2) $1<D\leq 2$:
The assertion follows from the fact that for all $l\geq 1$
\bee
\lim_{k\to\infty}\Lambda_l(1,k)=0  .
\ee
To see this, use
\bee 
\ln\prod_{j=i}^k\lambda_l(\beta_j)\leq 
\sum_{j=i}^k(\lambda_l(\beta_j)-1)
=-\sum_{j=i}^k \bigl[{l(l+N-2)\over 2 \beta_j}+O(\beta_j^{-2})\bigr] .
\ee
According to Prop. 2.1 the first sum diverges to $-\infty$; more
precisely, it behaves like $-(k)^{2-D}$ for $1<D<2$ and like $\ln k$
for $D=2$. The sum of the correction terms diverges for $1<D<3/2$ at
most like $k^{3-2D}$, for $D=3/2$ like $\ln k$ and converges for $D>3/2$.
In any case the asymptotic behavior for large $k$ is determined by the
leading term; so the product of the eigenvalues diverges to zero and the 
assertion follows.
\\
(3) $D>2$:
The product of the eigenvalues $\Lambda_l(k,\infty)$ converges absolutely 
to a nonzero value because
\bee
\sum_{j=i}^k|(\lambda_l(\beta_j)-1)|
\ee
converges for $k\to\infty$.
\endproof

One and two point functions simplify in the thermodynamic limit, provided
either $1\leq D\leq 2$ or we have free b.c. at 1 (i.e. $a=1$. Free b.c. at 
1 seem most natural anyway from the point of view of the $D$ dimensional 
lattice; if we consider $O(N)$ invariant observables, however, it is just 
as legitimate to choose $a\neq 0$, since in the thermodynamic limit this 
only means fixing the global $O(N)$ invariance.

We first discuss the case $1\leq D\leq 2$ and $a=0$. In this case the 
Hilbert space vector $\psi$ in the expression (\ref{1ptfct}) becomes 
proportional to the `ground state' $\psi_o$, which is the function 
identically equal to $1$ on $S_{N-1}$. $\psi_0$ spans the range of ${\cal 
P}_0$; it is invariant under all ${\cal T}_j$ and therefore we obtain
\bee
\la s_a(i)\ra_L={(\psi_0,s_a{\cal T}_{iL} \tilde\psi)
\over (\psi_0,\tilde\psi)}
\ee
and
\bee
\la s_a(i) s_b(k)\ra_L={(\psi_0,s_a {\cal T}_{ik} s_b
{\cal T}_{kL}\tilde\psi)
\over (\psi_0,\tilde\psi)}
\ee
Next we use the fact that $s_a$ maps the subspace ${\cal H}_0$ into
the subspace ${\cal H}_1$, to conclude (assuming $i\le k$)
\bee
\la s_a(i)\ra_L={(\psi_0,s_a \tilde\psi)
\over  (\psi_0,\tilde\psi)} \Lambda_1(i,L)
\ee
and
\bee
\la s_a(i) s_b(k)\ra_L={(\psi_0,s_a s_b{\cal T}_{kL}\tilde\psi)
\over  (\psi_0,\tilde\psi)}\Lambda_1(i,k).
\ee
Now we take the thermodynamic limit, using Proposition 3.1. We obtain
for $1\leq D\leq 2$:
\bee
\la s_a(i)\ra_\infty=0
\ee
and
\bee
\la s_a(i) s_b(k)\ra_\infty=(\psi_0,s_a s_b \psi_0)
\Lambda_1(i,k)
\ee
By $O(N)$ invariance it is easy to see that
\bee
(\psi_0,s_as_b\psi_0)={1\over N}\delta_{ab}
\ee
More generally, by the same reasoning, if $F$ is a bounded measurable 
function of finitely many spins $s(i_1), ...,s(i_k) $, and $\bar F$ its 
$O(N)$ average, 
\bee
\la F\ra_\infty=\la \bar F\ra_\infty.
\ee

Finally let us discuss what happens for $1\leq D\leq 2$ and $a\neq 0$, 
provided we consider $ON)$ invariant observables. So let $\bar F$ now be 
an $O(N)$ invariant bounded measurable function of finitely many spins.
In this case the Hilbert space vector $\psi$ in the expression 
(\ref{1ptfct}) becomes a function $\psi(a)$ with $(\psi(a),\psi_0)\neq 0$. 
But in the thermodynamic limit, again using Propositionn 3.1, we obtain
\bee
\la \bar F\ra_{\infty,a,b}=
{(\psi(a),\bar F\psi_0)\over (\psi(a),\psi_0)}=(\psi_0,\bar F\psi_0),
\ee
which is independent of $a$ and $b$.

We summarize what we have found for `low' dimension in

{\bf Theorem 3.2:} For $1\leq D\leq 2$ and $a=0$ (free b.c.~at 1) in the
thermodynamic limit, irrespective of the b.c. at $L$ (i.e. $b$) the 
following holds:
\begin{itemize}
\item $\la s_a(i)\ra_\infty=0$
\item $\la s_a(i) s_b(k)\ra_\infty={1\over N} \delta_{ab}
\prod_{j=i}^{k-1} \lambda_1(\beta_j)=
{1\over N} \delta_{ab}\Lambda_1(i,k) .$
\end{itemize}
In addition for any bounded measurable function $F$ of finitely many 
spins, and $\bar F$ its $O(N)$ average, we have
\bee
\la F\ra_\infty=\la\bar F\ra_\infty.
\ee
Furthermore for arbitrary $a$ \ \ 
$\la\bar F \ra_\infty$ is independent of $a$ and $b$.
\noindent
Finally for $k\to\infty$ we have the asymptotic behavior
\bee
\la s_a(i) s_b(k)\ra_\infty\sim 
\delta_{ab}{1\over N}\exp\left(-{N-1\over 2D2^D\beta(2-D)}k^{2-D}\right)
\ \ (1\leq D<2)
\ee
and
\bee
\la s_a(i) s_b(k)\ra_\infty\sim 
\delta_{ab}{1\over N}\exp\left(-{N-1\over 16\beta}\ln k\right)
\ \ (D=2).
\ee
\vskip15mm
For $D>2$ the situation is different and a little more involved.
We have
\bee
\la s_a(i)\ra_\infty={(\psi_0,s_a\tilde\psi)\over (\psi_0,\tilde\psi)}
\Lambda_1(i,\infty)
\ee
and
\bee
\la s_a(i) s_b(k)\ra_\infty={(\psi_0,s_a s_b {\cal T}_{k\infty}\tilde\psi)
\over (\psi_0,\tilde\psi)}
\Lambda_1(i,k).
\ee
If we choose Dirichlet b.c. by sending $\tilde\psi$ to a delta function
$\delta_{\e_N}$
concentrated at $\s=\e_N$, the one point function simplifies to
\bee
\la s_a(i)\ra_{\infty,{\rm Dir}}
=\delta_{aN}\Lambda_1(i,\infty)
\label{1pt}
\ee

We can analyze the 2 point function further by using the fact that only
the subspaces ${\cal H}_0$ and  ${\cal H}_2$ contribute here; generally
we have
\bee
(\psi_0,s_as_b\phi)={\delta{ab}\over N}
(\psi_0,\s^2{\cal P}_0\phi)
+(\psi_0,(s_as_b-{\delta_{ab}\over N}\s^2){\cal P}_2\phi)
\ee
for any $\phi\in{\cal H}$.
Inserting $\phi={\cal T}_{k\infty}\tilde\psi$ we obtain
\bee
\la s_a(i) s_b(k)\ra_\infty=
\delta_{ab}{1\over N}{(\psi_0,\s^2 \tilde\psi)\over
(\psi_0,\tilde\psi)}\Lambda_1(i,k)+
{(\psi_0,(s_as_b-\delta_{ab}{1\over N}\s^2)\tilde\psi)\over
(\psi_0,\tilde\psi)}\Lambda_1(i,k)
\Lambda_2(k,\infty).
\label{2ptdir}
\ee
Using the fact that
\bee
{(\psi_0,\s^2 \tilde\psi)\over(\psi_0,\tilde\psi)}=1
\ee
and taking the limit $k\to\infty$ we obtain
\bee
\lim_{k\to\infty} \la s_a(i) s_b(k)\ra_\infty=
\left[\delta_{ab}{1\over N}+
(\psi_0,(s_as_b-\delta_{ab}{1\over N}\s^2)\tilde\psi)\right]
\Lambda_1(i,\infty).
\ee
This latter expression is in general nonzero, but for Dirichlet b.c.
($\tilde\psi=\delta_{\e_N}$ we can evaluate it further, because in this
case
\bee
(\psi_0,(s_as_b-{\delta_{ab}\over N}\s^2)\tilde\psi)
=\delta_{aN}\delta_{b1}-\delta_{ab}{1\over N}
\ee
and we obtain
\bee
\lim_{k\to\infty} \la s_a(i) s_b(k)\ra_{\infty,{\rm Dir}}
=\delta_{aN}\delta_{bN}\Lambda_1(i,\infty).
\ee
Comparing with the result (\ref{1pt}) for the one point function we see
that the latter limit equals
\bee
\lim_{k\to\infty}\la s_a(i)\ra_{\infty,{\rm Dir}} 
\la s_a(k)\ra_{\infty,{\rm Dir}}
\ee
In other words, for Dirichlet b.c. at $L\to\infty$, the truncated two
point function
\bee
\la s_a(i);s_b(k)\ra_{\infty,{\rm Dir}}\equiv
\la s_a(i)s_b(k)\ra_{\infty,{\rm Dir}}-\la s_a(i)\ra_{\infty,{\rm Dir}}
\la s_b(k)\ra_{\infty,{\rm Dir}}
\ee
goes to 0 for $k\to\infty$. In this sense we may interprete the state 
with Dirichlet b.c. at $\infty$ as a pure phase of the system. Note also 
that by (\ref{asympt}) the limit is approached like $k^{2-D}$.

We can still work out the truncated two point function for Dirichlet b.c.
at $\infty$ in a little more detail: from eq. (\ref{2ptdir}) we have
\bea
\la s_a(i);s_b(k)\ra_{\infty,{\rm Dir}}
&=&\delta_{ab}{1\over N} \Lambda_1(i,k)
+\left(\delta_{aN}\delta_{b_N}-\delta_{ab}{1\over N}\right)
\Lambda_1(i,k)\Lambda_2(k,\infty)\cr
&-&\delta_{aN}\delta_{b_N}
\Lambda_1(i,\infty)\Lambda_1(k,\infty)
\eea
which can be rewritten as
\bee
\left(\delta_{aN}\delta_{b_N}-\delta_{ab}{1\over N}\right)
\Lambda_1(i,k)(\Lambda_2(k,\infty)-1)
+\delta_{aN}\delta_{b_N}\left(\Lambda_1(i,k)-
\Lambda_1(i,\infty)\Lambda_1(k,\infty)\right).
\ee
For $a=b\neq N$ this is manifestly positive. 
Furthermore, it is also straightforward to see that the $O(N)$ invariant 
truncated two point function is positive:
\bee
\la \s(i);\s(k)\ra>0.
\ee
where the semicolon is meant to include the dot symbolizing the scalar
product in $\R^N$.

Let us now summarize what has been found for $D>2$ in

{\bf Theorem 3.3:} For $D>2$ and and $a=0$ (free b.c. at 1) in the
thermodynamic limit
\begin{itemize}
\item
the one-point function is
\bee
\la s_a(i)\ra_\infty=\Lambda_1(i,\infty)
{(\psi_0,s_a\tilde\psi)\over(\psi_0,\tilde\psi)}
\ee
which is nonzero provided the scalar product in the numerator does not
vanish. This is in particular the case if $s_a=s_N$ and the b.c. 
parameter $b>0$.
\item
$\lim_{k\to\infty} \la s_a(i) s_b(k)\ra_\infty$ is generically nonzero.
\item For Dirichlet b.c. ($b\to\infty$) the one point function is
\bee
\la s_a(i)\ra_{\infty,{\rm Dir}}=\delta_{aN}\Lambda_1(i,\infty);
\label{1ptrepeat}
\ee
furthermore the two point function has the cluster property
\bee
\lim_{k\to\infty}\la s_a(i);s_b(k)\ra_{\infty,{\rm Dir}}=0.
\ee
\end{itemize}
The truncated two point function behaves like
\bee
\la s_a(i);s_a(k)\ra\equiv \la s_a(i) s_b(k)\ra_{\infty,{\rm Dir}}-
\la s_a(i)\ra_{\infty,{\rm Dir}} \la s_a(k)\ra_{\infty,{\rm Dir}}=
O(k^{2-D}).
\ee

From the two theorems we can easily derive results about the 
magnetization:

{\bf Cor.~3.4:} For $1\leq D\leq 2$ the spontaneous magnetization
$M_\infty=\lim_{L\to\infty} M_L$ vanishes, irrespective of the b.c.. 
For $D>2$ the model has spontaneous magnetization with suitable b.c.; for
Dirichlet b.c. $M_\infty=1$. 

{\it Proof:} The spontaneous magnetization $M_\infty$ is, according to
\ref{magn} a certain weighted average of the 1-point function 
$\la s_N(i)\ra_\infty$. The one point function converges for $i\to\infty$
to
\bee
\la s_N(\infty)\ra_\infty={(\psi_0,s_N\tilde\psi)\over (\psi,\tilde\psi)}
\ee
Any average of a convergent sequence is equal to its limit, so we have
\bee
M_\infty=\la s_N(\infty)\ra_\infty
\ee
For Dirichlet b.c., because $\tilde\psi$ becomes the distribution
$\delta_{\e_N}$ concentrated at $\e_N$, the last quantity is equal 
to 1.
\endproof

The physical interpretation of the fact that the truncated two point 
function $\la s_a(i)s_a(k)\ra$ decays faster than any power of $k$
for $1\leq D\leq 2$ is that the model is in its high temperature phase
for any value of $\beta$. Likewise the power-like decay in $D=2$
means that the model is critical for any $\beta$. The power-like decay
of $\la s_a(i);s_a(k)\ra$ for $D>2$ expresses the presence of a
Goldstone-like mode in the system.

The susceptibility is a little pathological in our model. As defined
above, it diverges in general (except for $D=1$) due to the freezing of 
the spins within one layer. Namely, if the truncated two point function
is nonnegative (as it is, according to the discussion above), we have
\bee
\chi_L={\beta\over V_L} \sum_{i,k=1}^L g_i g_k \la s_N(i);s_N(k)\ra
\geq \sum_{i=1}^L g_i^2 \la s_N(i)^2\ra={\beta\over N V_L}
\sum_{i=1}^L g_i^2
\ee
which diverges for $L\to\infty$.\\

\subsection{Presence and absence of asymptotic freedom}

In this subsection we discuss the issue of asymptotic freedom using a
definition of the Callan-Symanzik $\beta$-function $\beta_{CS}$ due to 
Patrascioiu \cite{Pprl54}. We use the truncated two point function
\bee
G_\beta(i,k)\equiv\la \s(i);\s(k)\ra
\ee
to define the following Renormalization Group (RG) invariant quantity:
\bee
F_\beta(i,k)\equiv {G_\beta(2i,2k)\over G_\beta(i,k)}
\ee
We rescale the lattice distances in $F_\beta$ and ask how this can be
compensated by a change of the coupling constant $g\equiv 1/\sqrt\beta$.

This compensation cannot be made exact by only changing $\beta$,
but it works aymptotically in the limit of many iterations, or
equivalently for the large distance asymptotics.

Since we have the asymptotic behavior (see (\ref{asympt}))
\bee
\la \s(i)\cdot\s(k)\ra\sim\exp\left({N-1\over 2^DD(D-2)\beta}
\left(i^{2-D}-k^{2-D}\right)\right)
\ee
(for $k>i$),
it is seen straightforwardly that a doubling of $i$ and $k$ can be
compensated by replacing $\beta$ with $2^{2-D}\beta$ or $g$ with
$2^{D/2-1}$. Of course there are corrections of order $1/\beta^k$ to
the exponent and we will compute the first one of these corrections in the
next section, but they do not affect the qualitative conclusions.

If we interpolate $F_\beta$ to obtain a smooth function on $\R^2$
we can do this rescaling infinitesimally and find that $F_\beta$
obeys, at least asympotically, the following RG equation:
\bee
\left(\partial_t+\beta_{CS}(g(t))\partial_g\right)F_\beta(e^ti,e^tk)=0
\label{RG}
\ee
This equation should really be interpreted at the definition of the
Callan-Symanzik $\beta$-function, i.e. we have to set
\bee
\beta_{CS}(g(t))=-{\partial_tF_\beta(e^ti,e^tk)\over
\partial_gF_\beta(e^ti,e^tk)}
\label{betaCS}
\ee
which yields, putting $t=0$ and $g(0)=g$,
\bee
\beta_{CS}(g)={D-2\over 2} g +O(g^2).
\ee
This result shows that for $D<2$ the model is asymptotically free,
whereas for $D=2$ it is critical for any $\beta>0$, i.e. we have a 
half-line of fixed points. This is true for any $N>1$, i.e. there is no
qualitative difference between the abelian and the nonabelian versions
of the model.

For $D>2$ the analysis is a little different, because we have to
take the truncation into account. The asymptotic behavior of the
truncated two point function is

\bea
\la \s(i);\s(k)\ra&\sim&\exp\left(-{N-1\over 2^DD(D-2)\beta}
\left(i^{2-D}-k^{2-D}\right)\right)\cr
&-&\exp\left(-{N-1\over 2^DD(D-2)\beta}\left(i^{2-D}+k^{2-D}\right)\right)
\eea
i.e.
\bee
\la \s(i);\s(k)\ra\sim -2{N-1\over 2^DD(D-2)\beta}k^{2-D}
\ee
Forming the renormalization group invariant $F_\beta$ we see that the
dependence on $\beta$ as well as the scale parameter drops out. The RG
equation is satisfied with
\bee
\beta_{CS}=0
\ee
Of course, strictly speaking, $\beta_{CS}$ is left undetermined by eq.
(\ref{RG}) and eq. (\ref{betaCS}). But $\beta_{CS}=0$ also is the
right answer if we consider the more general RG equation satisfied
by the truncated two point function itself:
\bee
\left(\partial_t+\beta_{CS}(g(t))\partial_g-(D-2)\right)
G_\beta(e^ti,e^tk)=0
\label{RG2}
\ee
We conclude that the model is thus also critical for $D>2$; the reason is
of course the presence of the Goldstone-like mode.

\subsection{Gibbs states and phase structure}
In this subsection we will discuss the set of Gibbs states of our model 
for the semi-infinite chain, obtained by taking the thermodynamic limit 
of finite chains with various b.c.. As before, we will use free b.c. 
at 1 ($a=1$), but the following should be noted: 

{\it Remark:} If a Gibbs state $\la .\ra$ is $O(N)$ invariant, we can
modify it by introducing an arbitrary measure for a particular spin,
without affecting the expectation values of $O(N)$ invariant observables.
In particular these expectation values become independent of the
measure $\psi$ chosen for the b.c. at 1.

Gibbs (= equilibrium) states can be characterized by the DLR 
equations \cite{DLR,BS}.
Consider a local observable ${\cal O}(\s(i),...\s(k))$, i.e. a
continuous function of a finite number of spins ($1\leq i<k$); then the
DLR equations imply that for any integer $r\geq 0$
\bee
\la {\cal O}\ra=\la\tilde{\cal O}_r\ra
\ee
where $\tilde{\cal O}_r$ is a function of $s(i)$ and $s(k+r)$
\bee
\tilde{\cal O}_r(\s,\s^\prime)={1\over Z}\int {\cal O}(\s(i),...\s(k))
\prod_{j=i}^{k+r-1}{\cal T}_j(\s(j),s(j+1)\prod_{j=i}^{k+r-1}d\nu(\s(j))
\ee
with
\bee
Z=\int \prod_{j=i}^{k+r-1}{\cal T}_j(\s(j),s(j+1)
\prod_{j=i}^{k+r-1}d\nu(\s(j))
\ee
So among other things, going from ${\cal O}$ to $\tilde{\cal O}_r$
attaches a product of $r$ transfer matrices to the observable.
We can now ask what happens if we send $r\to\infty$. In low dimensions 
($1\leq D\leq 2$) that product converges in the strong topology to
${\cal P}_0$, as we saw. This is already sufficient (by letting the adjoint
of those operators act `to the left') to conclude that for any Gibbs 
state and $1\leq D\leq 2$
\bee
\la {\cal O}\ra=\la{\cal O}\tilde{\cal P}_0\ra
\ee
where we use same symbol ${\cal P}_0$ also for the integral kernel of
the operator, i.e.
\bee
{\cal P}_0(\s,\s^\prime)=\psi_0(\s)\psi_0(\s^\prime)\equiv 1
\ee
The insertion of ${\cal P}_0$ has the same effect as free b.c.
at a point $m$ with $m>k$, so that we conclude
\bee
\la {\cal O}\ra=\la {\cal O}\ra_{\rm free}.
\ee

So remembering that we made the general assumption $a=1$ in this 
section, we have obtained the result:

{\bf Theorem 3.5:} For $1\leq D\leq 2$ with free b.c. at 1,  there is a 
unique Gibbs state obtained as a thermodynamic limit. In particular this 
Gibbs state is $O(N)$ invariant.

{\it Remark:} As noted above, this implies that for $O(N)$ invariant
observables, all b.c. at $i=1$ are equivalent.

Now we turn to the cae $D>2$. This turns out to be a little trickier
and we do not obtain a complete rigorous description of all the Gibbs
states. There is, however, a very natural conjecture.

Let us first state what we can prove:

{\bf Theorem 3.6:} For $D>2$ there is an uncountably infinite set of
Gibbs states parametrized by the set of probability measures on the
sphere $S_{N-1}$. The extremal states in this set are given by the 
probability measures $\delta_e$ concentrated in one point of 
$\e\in S _{N-1}$.\\
{\it Proof:} It follows from Proposition 3.1, as in the proof of
Theorem 3.3,
that the thermodynamic limit with a boundary measure $\tilde\psi$ exists
for any local observable. Theorem 3.3 says first of all that the one
point function in general will be different for different choices of
$\tilde\psi$. We can easily generalize the Dirichlet b.c. discussed
there by choosing
\bee
\tilde\psi=\delta_{\e}
\ee
where $\delta_{\e}$ is the delta function concentrated on a general
point $\e\in S_{N-1}$. The one point function will then be
\bee
\la \s(i)\ra_{\infty,\e}=\Lambda_1(i,\infty)\ \e
\ee
so all these states are different for different choices of $\e$.

It is also easy to see that these generalized Dirichlet states are all
extremal in the set of states given by boundary measures  $\tilde\psi$,
because they satisfy
\bee
\la \s(\infty)\cdot \e\ra_{\infty,\e}=1
\ee
whereas in any other state one has
\bee
\la \s(\infty)\cdot \e\ra< 1
\ee
\endproof

What we do not know rigorously is whether we have exhausted all the Gibbs
states by our b.c. $\tilde\psi$, but it is very plausible that this is so.
So we state

{\bf Conjecture 3.7:}
For $D>2$ the set of Gibbs states is given by the set of measures 
$\tilde\psi$ on the sphere $S_{N-1}$.

    %Nonperturbative Results
\section{Perturbation Theory}
\setcounter{equation}{0}
\subsection{Preliminaries}

Perturbation Theory (PT) is nothing but the application of Laplace's
method to the Gibbs measure. For a finite system, the Gibbs factor
is very sharply peaked around the ground state configuration(s). To
make PT work, we need a unique such ground state, and we achieve that
by a suitable choice of b.c.. The Gibbs measure is then to lowest order
(`tree level') approximated by a Gaussian centered at the ground state,
and a sequence of corrections to the Gaussian arises naturally by
Laplace's method.

For a finite system one can show easily that the resulting expansion
in inverse powers of $\beta$ is asymptotic to the true expectation values.
The usual formal PT procedure takes the thermodynamic limit term by term
and hopes, if that limit exists, to obtain an expansion that is asymptotic
to the infinite volume Gibbs state. It is well known that this hope fails
in some cases (see \cite{BR,Pprl54,PSsi}, and here we will find that it
fails in general for dimension $D\leq 2$.

From now on we assume that we have b.c. characterized by $b>0$ and
$a\geq 0$. Then the ground state configuration is unique and is described
by $\s(j)=\e_N$ for all $j$. For large $\beta$ with high probability the
spins will fluctuate not very far from $\e_N$. This motivates the
introduction of Cartesian coordinates on the sphere $S_{N-1}$ (as in the 
classic paper of Br\'ezin and Zinn-Justin \cite{BZJ}) to describe these
fluctuations:
\bee
\s(j)=\left(\begin{array}{c}
\beta^{-\frac{1}{2}}\P_j \\
\sigma_j
\end{array}\right), \qquad \mbox{with} \quad
\sigma_j:=\pm (1-\beta^{-1}\P_j^2)^{\frac{1}{2}};
\ee
These coordinates are singular at the equator, but since we are interested
in an asymptotic expansion in powers of $1/\beta$, we can ignore this
fact. We can actually limit ourselves to integrating over the upper
hemisphere, i.e. always choose the + sign in the definition of $\sigma_j$.
Finally we can extend the integration over each $\pi_j$ to all of
$\R^{N-1}$. All these changes have only an exponentially small effect (in 
$\beta$) on the integrals, as long as we work with a finite system 
$L<\infty$, and therefore they do not affect PT.

The partition function can therefore be replaced by
\bea
&&Z_L(h,a,b,D)=\cr&&\int
\exp\biggl[\sum_{j=1}^{L-1}b_j \left(\P_j\cdot\P_{j+1}
+\beta\left(1-\beta^{-1}\P_j^2\right)^{1\over 2}
\left(1-\beta^{-1}\P_{j+1}^2\right)^{1\over 2}\right)\cr
&+&\beta h \sum_{j=1}^L g_j
\left(1-\beta^{-1}\P_j^2\right)^{1\over 2}
+\beta a \left(1-\beta^{-1}\P_1^2\right)^{1\over 2}
+\beta b\left(1-\beta^{-1}\P_L^2\right)^{1\over 2}
\biggr]\cr
&\times&\exp\left[-{1\over 2}\sum_{j=1}^L
\ln\left(1-\beta^{-1}\P_j^2\right)\right]
\prod_{j=1}^Ld\P_j.
\eea
and accordingly for the expectation values. In other words, for the 
purpose of PT we are reduced to studying the Gibbs measures
\bee
d\mu_L(\P_1,...,\P_L)={1\over Z_L}\exp(A_L)\prod_{j=1}^Ld\P_j
\ee
with
\bea
&&A_L=
\biggl[\sum_{j=1}^{L-1}b_j \left(\P_j\cdot\P_{j+1}
+\beta\left(1-\beta^{-1}\P_j^2\right)^{1\over 2}
\left(1-\beta^{-1}\P_{j+1}^2\right)^{1\over 2}\right)\cr
&+&\beta h \sum_{j=1}^L g_j
\left(1-\beta^{-1}\P_j^2\right)^{1\over 2}
+\beta a \left(1-\beta^{-1}\P_1^2\right)^{1\over 2}
+\beta b\left(1-\beta^{-1}\P_L^2\right)^{1\over 2}
\biggr]\cr
&-&{1\over 2}\sum_{j=1}^L
\ln\left(1-\beta^{-1}\P_j^2\right)
\eea
It is now clear how to proceed; we expand $A$ in powers of $\beta$; in
this paper we will be content to do this to order $1/\beta$:
\bee
A=2^{D-1}\beta V_L(1+h)+a\beta b_1+b\beta +A^{(0)}+A^{(1)}+O(\beta^{-2})
\ee
where the piece $O(\beta^0)$ is quadratic in the $\P_j$ variables:
\bee
A^{(0)}=-{1\over 2}\sum_{j=1}^{L-1}b_j(\P_{j+1}-\P_j)^2-
{h\over 2}\sum_{j=1}^{L}
g_j\P_j^2-{a\over 2}\P_1^2-{b\over 2}\P_L^2
\ee
and the term $O(\beta^{-1})$ is
\bea
A^{(1)}&=&-{1\over 2\beta}\biggl[\sum_{j=1}^{L-1}b_j
\left({\P_{j+1}^2-\P_j^2\over 2}\right)^2\cr
&+&{h\over 4}\sum_{j=1}^Lg_j(\P_j^2)^2-\sum_{j=1}^L\P_j^2
+{1\over 4} a(\P_1^2)^2+{1\over 4}b(\P_L^2)^2\biggr].
\label{action1}
\eea
As long as not $a=b=h=0$, $-A^{(0)}$ is a positive definite quadratic
form in the $\P_j$ which we write as
\bee
A^{(0)}\equiv -{1\over 2}(\P,Q\P)
\ee
where $\P$ stands for the vector
$\left(\pi_i^a\right)_{i=1,...,L}^{a=1,...,N-1}$ in $\R^{(N-1)L}$ and
$Q$ has matrix elements $Q_{ik}^{ab}=\delta^{ab}q_{ik}$.
We combine $A^{(0)}$ with the Legesgue measure to produce the Gaussian
probability measure
\bee
d\mu_C(\P)={1\over Z_L^{(0)}}\exp(-{1\over 2}(\P,Q\P)
\ee
with covariance $C=Q^{-1}$. $C$ is proportional to the identity operator
in the internal space $\R^{N-1}$ , i.e. its matrix elements are of the
form $C_{ik}^{ab}=\delta^{ab}c_{ik}$. $A^{(1)}$, even though it contains 
also terms quadratic in $\P$, is treated as an interaction because it 
is of order $\beta^{-1}$.

\subsection{Spin two point function}

We first derive the explicit form of PT up to order $\beta^{-2}$ for
the invariant spin-spin correlation in terms of the covariance of the
Gaussian measure up to order $\beta^{-2}$. First note that
\bee
\la \s(i)\cdot\s(k)\ra=1-{1\over 2\beta}\la(\P_i-\P_k)^2\ra
-{1\over 8\beta^2}\la \left(\P_k^2-\P_k^2\right)^2\ra+O(\beta^{-2})
\ee
provided we have $a,b,h\geq 0$ and at least one of them $>0$, because 
then the ground state will have all spins aligned parallel to $\e_N$.
As usual, PT is generated by expanding also the interaction in
inverse powers of $\beta$ in the Gibbs measure; as is well known, the
correct normalization leads to the phenomenon that `vacuum graphs cancel',
i.e.~only terms survive in which the interaction is connected by
`lines' (covariances) to the observable. Thus
\bea
&&\la \s(i)\cdot\s(k)\ra=1-{1\over 2\beta}\la(\P_i-\P_k)^2\ra_c
-{1\over 8\beta^2}\la\left(\P_i^2-\P_k^2\right)^2\ra_c \cr
&-&\la {1\over 2\beta}(\P_i-\P_k)^2; A^{(1)}\ra_c+O(\beta^{-3})
\eea
where the semicolon indicates that only contributions are to be taken
that connect the expression to the left with that to the right of it.
$\la.\ra_c$ denotes the Gaussian expectation value with covariance $c$
and $A^{(1)}$ is defined in eq.(\ref{action1}).

The term order $\beta^{-1}$ is therefore
\bee
\la \s(i)\cdot\s(k)\ra^{(1)}=-{N-1\over 2\beta}(c_{ii}+c_{kk}-2c_{ik})
\ee
To order $\beta^{-2}$ we find
\bee
\la \s(i)\cdot\s(k)\ra^{(2)}=(I)+(II)+(III)+(IV)+(V)
\ee
with
\bee
(I)\equiv
-{1\over 8\beta^2}\la\left(\P_i^2-\P_k^2\right)^2\ra_c=
-{N-1\over 8\beta^2}
\biggl[(c_{ii}^2+c_{kk}^2)(N+1)-4c_{ik}^2
-2(N-1)c_{ii}c_{kk} \biggr],
\ee

\bee
(II)\equiv-{1\over 4\beta^2}\sum_{j=1}^L\la\left(\P_i-\P_k\right)^2;
\P_j^2\ra_c
=-{N-1\over 2\beta^2}\sum_{j=1}^L(c_{ij}-c_{kj})^2,
\ee
\bea
(III)&\equiv&{1\over 16\beta^2} \sum_{j=1}^{L-1}b_j
\la(\P_i-\P_k)^2;\left(\P_j^2-\P_{j+1}^2\right)^2\ra_c\cr
&=&{N-1\over 4\beta^2}\sum_{j=1}^{L-1}b_j
\Biggl\{(N+1)\biggl[(c_{jj}(c_{ij}-c_{kj})^2+c_{j+1,j+1}(c_{i,j+1}
-c_{k,j+1})^2\biggr]
\cr
&-&(N-1)\biggl[c_{jj}(c_{i,j+1}-c_{k,j+1})^2+
c_{j+1,j+1}(c_{ij}-c_{kj})^2\biggr]
\cr
&-& 4c_{j,j+1}(c_{ij}-c_{kj})(c_{i,j+1}-c_{k,j+1})\Biggr\},
\eea
\bee
(IV)\equiv
{h\over 8\beta^2}\sum_{j=1}^Lg_j\la\left(\P_i-\P_k\right)^2;
\left(\P_j\right)^2 \ra
={(N-1)(N+1)\over 4\beta^2}h\sum_{j=1}^Lg_j c_{jj}(c_{ij}-c_{kj})^2
\ee
and finally the boundary contribution
\bea
(V)&\equiv&
{1\over 8\beta^2}\la\left(\P_i-\P_k\right)^2;
a\left(\P_1^2\right)^2+b\left(\P_L^2\right)^2 \ra \cr
&=&{(N-1)(N+1)\over 8\beta^2}\left[ac_{11}(c_{i1}-c_{k1})^2
+bc_{LL}(c_{iL}-c_{kL})^2\right]
\eea
\noindent
It is advantageous to rewrite contribution $(III)$ as follows:
\bee
(III)=(IIIa)+(IIIb)+(IIIc)
\ee
with
\bee
(IIIa)\equiv {(N-1)N\over 4\beta^2}\sum_{j=1}^{L-1} b_j
(c_{jj}-c_{j+1,j+1})
\left[(c_{ij}-c_{kj})^2-(c_{i,j+1}-c_{k,j+1})^2\right]
\ee
\bee
(IIIb)\equiv {N-1\over 4\beta^2}\sum_{j=1}^{L-1} b_j
(c_{jj}+c_{j+1,j+1})
(c_{ij}-c_{kj}-c_{i,j+1}+c_{k,j+1})^2
\ee
\bee
(IIIc)\equiv {N-1\over 2\beta^2}\sum_{j=1}^{L-1} b_j
(c_{jj}+c_{j+1,j+1}-2c_{j,j+1})
(c_{ij}-c_{kj})((c_{i,j+1}-c_{k,j+1})
\ee

\subsection{ The covariance without magnetic field}

To proceed, we have to compute the covariance $C$ more explicitly.
We only do this for the simplest case of free b.c. at 1 and no magnetic
field, i.e. $b=h=0$, but with $a>0$. We can ignore the internal space in 
this computation, i.e. put $N=2$. Then, with $\x\in\R^L$, we have
\bee
(\x,Q\x)=\sum_{j=1}^{L-1}b_j(x_{j+1}-
x_j)^2+ax_1^2
\ee
For convenience we define in the following $b_0\equiv a$ and $b_L\equiv 
b$. $Q$ is already given in the form
\bee
Q=L^TL
\ee
where $L$ is a lower triangular (actually bidiagonal) matrix with 
elements
\bea
l_{kk}&=&\sqrt{b_{k-1}}\ \ \ {\rm and} \cr
l_{k+1,k}&=&-\sqrt{b_k}\ \ \ {\rm for} \ \ k=1,2,...,L
\eea
all other elements being zero. We now split
\bee
L=D+N
\ee
where $D={\rm diag}(\sqrt b_0,...\sqrt{b_{L-1}})$ and $N$ is nilpotent.
Then
\bee
C=Q^{-1}=(\1+D^{-1}N)^{-1}D^{-2}(\1+N^TD^{-1})^{-1}
\ee
$D^{-1}N$ has only nonzero matrix elements $d_{ik}$ for $i=k+1$, and they
are all equal to $-1$. Therefore $Y\equiv (\1+D^{-1}N)^{-1}$ has the 
matrix elements
\bea
y_{ik}&=&1\ \ \ {\rm for}\ \  i\geq k  \cr
y_{ik}&=&0\ \ \ {\rm otherwise}
\eea
So, using the shorthand $m_{ik}\equiv {\rm min}(i,k)$ we obtain
for the covariance
\bee
c_{ik}=\sum_{j=1}^L b_{j-1}^{-1}y_{ij}y_{kj}=\sum_{r=0}^{m_{ik}-1}b_j^{-1}
\ee
for $b\equiv b_L=0$. If we introduce the further shorthand
\bee
B_{ik}\equiv\sum_{j=i}^{k-1}b_j^{-1}
\ee
we can write the covariance we found as
\bee
c_{ik}(a,0)=B_{0m_{ik}}=B_{1m_{ik}}+a^{-1}
\label{cova0}
\ee
where the second argument is reserved for the parameter $b$.

The covariance for general values of $a$ and $b$ (still without a
magnetic field) is obtained from this by realizing that changing the
b.c. at $L$ is a rank one perturbation:
\bee
c_{ik}(a,b)=c_{ik}(a,0)-\left(b^{-1}+c_{LL}(a,0)\right)^{-1}
c_{iL}(a,0)c_{kL}(a,0).
\ee
This can be written in more compact form if we use the definitions
$M_{ik}\equiv {\rm max}(i,k)$ and 
\bee
B\equiv (a^{-1}+b^{-1}+B_{1L})^{-1}={1\over B_{0,L+1}}:
\ee
as
\bee
c_{ik}(a,b)=B B_{0m_{ik}}B_{M_{ik},L+1}
\label{covar}
\ee
We note some special cases:
\bee
c_{ik}(0,b)=B_{M_{ik}L}+b^{-1}=B_{M_{ik},L+1}
\label{cov0b}
\ee
\bee
c_{ik}(\infty,b)=B B_{1m_{ik}}B_{M_{ik},L+1}
\label{covinfb}
\ee
\bee
c_{ik}(a,\infty)=
B B_{M_{ik}L}B_{0m_{ik}}
\label{covainf}
\ee

We also note the following combination of covariances:
\bee
c_{ii}(a,b)+c_{kk}(a,b)-2c_{ik}(a,b)=B_{ik}(1-BB_{ik})
\label{e1}
\ee

The behavior for large $L$ and the asymptotic behavior of the infinite
volume covariances for $k\to\infty$ can be obtained from Proposition 2.1,
which implied that for $1\leq D<2$ $B_{ik}=O(k^{2-D})$, for $D=2$
$B_{ik}=O(\ln(k))$ and for $D>2$ $B_{ik}=O(1)$. The thermodynamic limit
is particularly easy to take for $b=0$, because in that case by eq.
(\ref{cova0}) the covariance does not show any dependence on $L$.

In general for $1\leq D\leq 2$
\bee
\lim_{L\to\infty} c_{ik}(a,b)=B_{0m_{ik}}
\label{limcovar}
\ee
independently of $b$, provided $a>0$. For  $1\leq D\leq 2$ and $a=0$
the thermodynamic limit of the covariance does not exist.

For $D>2$  the thermodynamic limit always exists, but it depends on the
b.c. parameter b.

For the general case with a magnetic field there are no such simple
expressions for the covariance. A very detailed discussion with many
explicit and lengthy formulae can be found in \cite{Y}.

\subsection{Explicit evaluation to one loop}

We will first discuss the simplest case of the `energy', that is
the two point function of two neighboring spins.

To order $\beta^{-1}$ (`tree level'), we have
\bee
\la \s(i)\cdot\s(i+1)\ra=1-{N-1\over 2\beta}(c_{ii}+c_{i+1,i+1}
-2c_{i,i+1})+O(\beta^{-2})
\ee
Using eq.(\ref{e1}) we thus find
\bee
\la \s(i)\cdot\s(i+1)\ra=1-{N-1\over 2\beta_i}\left(1-Bb_i^{-1}\right)
+O(\beta^{-2})
\ee
and taking the thermodynamic limit of the first order term we
obtain
\bee
\la \s(i)\cdot\s(i+1)\ra^{(1)}=b_i^{-1}\ \ \ (1\leq D\leq 2)
\ee
and
\bee
\la \s(i)\cdot\s(i+1)\ra^{(1)}=b_i^{-1}\left(1-
{b_i^{-1}\over b^{-1}+B_{0\infty}}\right)\ \ \ (D>2)
\ee
This means that for $D>2$ already at tree level PT shows a dependence on 
b.c. even for such a simple $O(N)$ invariant observable. By expanding the
nonperturbative solution of the model, analyzed in the previous section, 
one can see that this is a real effect, related to the occurrence of
spontaneous symmetry breaking (SSB). The dependence on the b.c. parameter
$b$ drops out, however, if we use free b.c. at 1 ($a=0$). This happens
because for an invariant observable we can always introduce an arbitrary 
probability measure for one and only one spin, without any effect, as 
explained earlier.

To the next order (`one loop level') the computation becomes a little
more involved. Since for $D>2$ already the tree level term depends on
the b.c., we assume from now on $1\leq D\leq 2$.

We first consider the case $a>0,\ \ b=0$. It is easy to see that
in this case term (V) vanishes in the thermodynamic limit. Term (IV)
is absent because we put the magnetic field $h$ equal to zero. 
For the other terms, after some trivial algebra one obtains
\bee
(I)=-{N-1\over 8\beta_i^2}\left[N+1+4b_iB_{0i}\right],
\ee
\bee
(II)=-{N-1\over 8\beta_i^2}\left[4(L-i)\right],
\ee
\bee
(III)={N-1\over 8\beta_i^2}\left[2(N-1)+4(L-i)+4b_iB_{0i}\right],
\ee
which adds up to
\bee
\la \s(i)\cdot\s(i+1)\ra^{(2)}(a,0)={(N-1)(N-3)\over 8\beta_i^2}
\label{encorrect}
\ee

The case $h=0$ with general b.c. can also be worked out explicitly. 
First note that the term (V) still does not contribute in the 
thermodynamic limit: we have
\bee
(V)={(N-1)(N+1)\over 8\beta^2}\left[ac_{11}(c_{i1}-c_{k1})^2
+bc_{LL}(c_{iL}-c_{kL})^2\right].
\ee
For $a>0$ and any $b$ the first term goes to 0 as $L\to\infty$, since
by eq.(\ref{limcovar}) $\lim_{L\to\infty}(c_{iL}-c_{kL})=0$. For the 
second term we notice
\bee
bc_{LL}(c_{iL}-c_{kL})^2=b B^3 B_{0m} B_{m,L+1}(B_{0i}-B_{0k})^2 b^{-2}=
b B^3 B_{0m} B_{m,L+1} B_{ik}^2\to 0
\ee
for $L\to\infty$.

Inserting our formula eq.(\ref{covar}) into the expressions at the end
of Subsection 4.2, the result for $(II)$ and $(III)$ splits into sums
from 1 to $i-1$ and from $i$ to $L-1$ or $L$. For the finite sums
the limit $L\to\infty$ can be taken termwise and is actually zero,
because each propagator carries a factor $B$ which goes to 0 as 
$L\to\infty$. After some algebra we are left with
\bee
(I)=-{N-1\over 8\beta_i^2}\left[N+1+4b_iB_{0i}\right],
\ee
\bee
(II)=-{N-1\over 8\beta_i^2}B^2\sum_{j=i+1}^L B_{j,L+1}^2
\ee
\bee
(IIIa)\sim{(N-1)N\over 4\beta^2} \left\{ b_i^{-2}+B^2 B_{ik}^2
\sum_{j=i+1}^{L-1}\left[2b_j^{-1}B_{j,L+1}-4BB_{j,L+1}^2\right]\right\}
\ee
\bee
(IIIb)\sim{N-1\over 4\beta^2} \left\{ b_i^{-2}-B^3 B_{ik}^2
\sum_{j=i+1}^{L-1}2b_j^{-1}B_{j,L+1}^2)\right\}
\ee
\bee
(IIIc)\sim{N-1\over 2\beta^2}B_{ik}^2\sum_{j=i+1}^{L-1} \left[
B^3 B_{j,L+1}^2-B^2 b_j^{-1} B_{j,L+1}-B^3 b_j^{-1} B_{j,L+1}^2
\right]
\ee
Here the symbol $\sim$ means equality up to terms vanishing in the limit
$L\to\infty$. In arriving at these expressions we used the fact that some 
terms vanish in the limit $L\to\infty$:

{\bf Lemma 4.1:}: For $1\leq D\leq 2$ the following holds in the limit
$L\to\infty$:\\
\bee
(1)\ \  B^3\sum_{j=i+1}^{L-1}b_j^{-2}B_{j,L+1}\to 0,\\
\ee
\bee
(2)\ \  B^3\sum_{j=i+1}^{L-1}b_j^{-2}B_{0j}\to 0,\\
\ee
\bee
(3)\ \  B^3\sum_{j=i+1}^{L-1}b_j^{-3}\to 0.
\ee

{\it Proof:} \\$0<(1)\leq B^2\sum_{j=i+1}^{L-1}b_j^{-2}$\\
$0<\ (2) \leq  B^2\sum_{j=i+1}^{L-1}b_j^{-2}$\\
Using Proposition 2.1, it is seen easily that both upper bounds go to 
zero for $L\to\infty$. (3) is even more elementary.  \endproof

There are also sums that converge to nonzero limits:

{\bf Lemma 4.2:} For $1\leq D\leq 2$ and $b>0$
\bee
(1)\ \ I_1\equiv
\lim_{L\to\infty} B^2\sum_{j=i+1}^{L-1}b_j^{-1}B_{j,L+1}={1\over 2}\\
\ee
\bee
(2)\ \ I_2\equiv
\lim_{L\to\infty} B^3\sum_{j=i+1}^{L-1}b_j^{-1}B_{j,L+1}^2={1\over 3}.
\ee
For $b=0$ $I_1$ and $I_2$ vanish, because each term in the sums vanishes
before taking the limit.

{\it Proof}: The two statements are proven by `summation by parts'.
First note that for any sequences $f_i,g_i$ we have
\bee
 \sum_{j=A}^B [f_{j+1}-f_j]g_j=-\sum_{j=A}^B[g_{j+1}-g_j]f_{j+1}
+f_{B+1}g_{B+1}-f_Ag_A.
\label{sbp}
\ee
Applying this to the sum in (1) above and noting that 
$b_j^{-1}=B_{j+1,L+1}-
B_{j,L+1}$, we obtain
\bee
\sum_{j=i+1}^{L-1}b_j^{-1}B_{j,L+1}=\sum_{j=i+1}^{L-1}b_j^{-2}+
{1\over 2}B_{i+1,L+1}^2-{1\over 2}b_L^{-2}.
\ee
From this the assertion (1) follows easily, using Lemma 4.1.

Applying the formula (\ref{sbp}) to the expression in (2) above yields,
after some simplification,
\bee
\sum_{j=i+1}^{L-1}b_j^{-1}B_{j,L+1}^2=\sum_{j=i+1}^{L-1}b_j^{-2}B_{j,L+1}
-{1\over 3}\sum_{j=i+1}^{L-1}b_j^{-3}-{1\over 3}b_L^{-3}
+{1\over 3}B_{i+1,L+1}^3.
\ee
Again the assertion follows from Lemma 4.1.
\endproof

Now we can add up all the contributions; the divergent terms cancel and we 
obtain
\bee
(I)+(II)+(III)\sim {N-1\over 8\beta_i^2}\left\{-4B^2B_{L,L+1}^2
+(N+1)-8(N+1)I_1+4NI_2\right\}
\ee

For $b=0$ $I_1$ and $I_2$ vanish, $B^2B_{L,L+1}^2=1$ and we recover our 
old result (\ref{encorrect}). For $b>0$, however, the term
$B^2B_{L,L+1}^2$ vanishes and the sums $I_1$ and $I_2$ converge to 1/2 
and 1/3.
So we obtain
\bee
\la \s(i)\cdot\s(i+1)\ra^{(2)}(a,b)={(N-1)(N-5)\over 24\beta_i^2}.
\ee
for $b>0$.

So we have found that in our model PT gives results depending on b.c.
for any dimension (except for $N=2$)! While this is a sensible
result for $D>2$ due to the occurrence of SSB, it signals a disease of PT 
for $1\leq D\leq 2$, where we have a unique Gibbs state. The right 
asymptotic expansion to order $\beta^{-2}$ is obtained only with free 
b.c., where the thermodynamic limit is reached already for finite $L$. 
Free b.c. are `DLR b.c.' for our model, corresponding to exactly  
integrating out the variables outside our box.

We now turn to the general case $\la \s(i)\cdot\s(k)\ra^{(2)}(a,b)$ for
$i>k$. The computation is similar, only this time we have to split the 
sums into 3 parts: from 1 to $i-1$, from $i$ to $k-1$ and from $k$ to $L$ 
or $L-1$. For the sums of finite range we can again take the termwise 
limit $L\to\infty$, which makes the sums from 1 to $i-1$ disappear. The 
sums from $i$ to $k-1$ produce, among others 
\bee
\sum_{j=i}^{k-1}b_j^{-1} B_{ij}=-{1\over 2}\sum_{j=i}^{k-1}b_j^{-2}
+{1\over 2}B_{ik}^2 .
\ee
Using this formula is is not difficult to find the final result:

{\bf Proposition  4.3:} The one loop PT result for the spin-spin two point
function is for free b.c. ($b=0$):
\bee
\la \s(i)\cdot\s(k)\ra^{(2)}(a,b)={(N-1)(N-3)\over 8\beta^2} B_{ik}^2
-2(N-1)\left[\sum_{j=i}^{j=k-1}b_j^{-2}-B_{ik}^2\right]
\ee
and represents the correct aymptotics in the thermodynamic limit. Taking
the thermodynamic limit termwise for $b>0$, we instead obtain the 
incorrect result
\bee
\la \s(i)\cdot\s(k)\ra^{(2)}(a,b)={(N-1)(N-5)\over 8\beta^2} B_{ik}^2
-2(N-1)\left[\sum_{j=i}^{j=k-1}b_j^{-2}-B_{ik}^2\right]
\ee

A final remark concerns free b.c. at site 1 ($a=0$): since the result
for $b=0$ is strictly independent of $L$ as well as $a$, and since
the labeling of the sites can be inverted ($j\to L-j$), the result
for $a>0,\ \ b=0$ is identical with the result for $a=0,\ \ b>0$, and
therefore produces the correct asymptotics.

Comparing the results obtained with different b.c, we can sum up what we 
have found as follows: the second order PT term for the invariant two 
point function, computed by the conventional termwise thermodynamic limit 
with b.c. parameters $a,b>0$ differs from the correct second order term 
by
\bee
\delta \la \s(i)\cdot \s(k)\ra^{(2)}(a,b)={(N-1)(N-2)\over 12\beta^2} 
B_{ik}^2 
\ee
This dependence on b.c. we have found highlights a general problem of
PT: an infrared regulator is needed, but the standard procedure of
removing that regulator termwise yields ambiguous results. In general they 
are incorrect, and only in our simplified model we are in the lucky 
situation to know the DLR b.c. which yield the true answer. 

The case $N=2$ is special: we have not found manifest signs of a
disease of PT. In fact for the translation invariant $O(2)$ model it has
been {\it proven} in \cite{bricmont} that standard PT yields the correct 
asymptotic expansion. 

\subsection{Remarks on the magnetic field as infrared regulator}

A popular infrared regulator in the non-linear $O(N)$ $\sigma$-models is 
the magnetic field. It was used for the first detailed study of 
the perturbative Renormalization Group and the asymptotic freedom 
predicted by it \cite{BZJ}. On the other hand it has been known for many 
years that the usual procedure of doing the perturbation expansion in the 
presence of a magnetic field and than removing it termwise yields 
incorrect results already for a finite number of spins \cite{Hasenfratz} 
and in $D=1$ \cite{BR}. Since our models are essentially one-dimensional, 
the same phenomenon is expected to occur also here. This is discussed in 
detail in \cite{Y}. 
 
Here it would lead us too far afield to go into this very technical 
matter, which involves interesting methods from the theory of continued 
fractions. But it should be seen that introducing the b.c. parameters $a$ 
and $b$ essentially amounts to the introduction of a {\it local} magnetic 
field, and we have seen in the previous subsection that for $N>2$ with 
$a>0$ and $b>0$ PT produces incorrect results, independent of the values 
of $a$ and $b$; hence sending $a,b\to 0$ in the end does not help. So it 
should be clear that one cannot hope for anything better with a global 
magnetic field.
    %Perturbation Theory
%\input magn.tex  %PT with magnetic field
\section{Percolation Properties}
\setcounter{equation}{0}

In this section we discuss briefly some percolation properties of the 
model. Even though percolation is rather trivial in our essentially 
one-dimensional systems, we find it worthwhile to check whether the 
general ideas of \cite{PSperc1,PSperc2,PSperc3} apply in this case. Not 
surprisingly, we find again the familiar dichotomy between the situation 
in $D\le 2$ and $D>2$. 

We are interested in the percolation properties of sets defined by the 
spin $\s$ lying in certain open connected subsets $A$ of the sphere 
$S_{N-1}$, such as the `polar caps' ${\cal P}_\epsilon^+\equiv \s\cdot 
\e_N>\epsilon/2$ and `equatorial strips' ${\cal S}_\epsilon\equiv |\s\cdot 
\e_N|<\epsilon/2$ discussed in \cite{PSperc1,PSperc2,PSperc3}. For 
simplicity we say `a certain subset $A\subset S_{N-1}$ percolates' when 
we mean that the set of points of our lattice $\{i\in\Z^D|\s(i)\in A\}$ 
percolates.  

Let us first discuss the case of no symmetry breaking ($1\le D\le 2$):

{\bf Theorem 5.1:} For $1\le D\le 2$ a subset $A\subset S_{N-1}$ whose 
complement is open and nonempty never percolates.

{\it Remark:} If one interpretes the model as living on $\Z^D$, this can 
be viewed for $D=1$ as the formation of `islands' and for $D=2$ as the 
`ring formation' discussed in \cite{PSperc1,PSperc2,PSperc3}.

{\it Proof:} Assume the contrary. Consider the characteristic function
$\chi_A(s(i))$. Then by Theorem 3.2
\bee   
\la \chi_A(s(i))\ra=\int_A d\nu(\s),
\ee
which is a number independent of $i$ and $<1$. On the other hand, if $A$ 
percolates,
\bee 
\lim_{i\to\infty} \la \chi_A(s(i))\ra\geq 
\la \lim_{i\to\infty}\chi_A(s(i)\ra =1
\ee
by Fatou's lemma, which is a contradiction. \endproof

We now turn to the case of spontaneous symmetry breaking ($D>2$). In this 
case it is to be expected that in the Gibbs state $\la . 
\ra_{\infty, {\rm Dir}}$ , obtained as the thermodynamic limit with 
$\e_N$-Dirichlet b.c., there is percolation of any neighborhood of the 
`north pole' $\e_N$. To actually prove this requires rather detailed 
technical estimates (cf. \cite{Y}). Here we will limit ourselves to giving 
a simple proof of this fact for $D>3$) and then show that it can be 
extended to $D>5/2$.

{\bf Theorem 5.2:} For $D>5/2$, in the state obtained as the 
thermodynamic limit with $\e_N$-Dirichlet b.c., any open set $A$ 
containing the point $e_N\in S_{N-1}$ percolates.

{\it Proof:} We will show that the probabilities for the events
\bee
a_i\equiv \left\{\s(i)\notin A\right\}
\ee
are summable; the theorem then follows from the Borel-Cantelli Lemma,
which states that in this case with probability 1 only finitely many of
the events $a_i$ occur.

We have the following
 
{\bf Proposition 5.3:} Let $\chi^\epsilon_i$ be the characteristic
function of the set $B\equiv\{\s(i)\in S_{N-1}|\ |\s(i)-\e_N|>\epsilon
\}$.
Then 
\bee
\la \chi^\epsilon_i \ra_{\infty,\rm Dir}<{c i^{2-D}}
\ee

{\it Proof:} Obviously for any $n>0$
\bee
\chi^\epsilon_i\le \left({|\s(i)-\e_N|\over \epsilon}\right)^n.
\label{tcheby}
\ee
Using $n=2$ and the results of Section 3 we obtain: 
\bea
\la |s(i)-\e_N|^2\ra_{\infty,{\rm Dir}} &=&\la 
\sum_{a=1}^{N-1}s_a(i)^2\ra_\infty
={N-1\over N}\left(1-{\Lambda_2(\infty)\over \Lambda_2(i)}y\right)\cr
&\sim&-{l(l+N-2)\over 2\beta}
{1\over 2^D D(D-2)} i^{2-D} .
\eea
The last expression is clearly summable over $i$ for $D>3$, so this proves 
percolation for $D>3$.

To extend the proof to $D>5/2$, we have to work a little more. Again by 
the Borel-Cantelli Lemma, the claim will follow from a sharpening of 
Proposition 5.3:

{\bf Proposition 5.4:} 
\bee
\la \chi^\epsilon_i \ra_{\rm Dir}<{c i^{2(2-D)}}
\ee

{\it Proof:} We choose  $n=4$ in the inequality \ref{tcheby}. Denoting
$\s(i)\cdot \e_N\equiv z$ we expand $|s(i)-\e_N|^4$ in Gegenbauer 
polynomials in $z$:
\bee
|s(i)-\e_N|^4=1-2z^2+z^4=a_o+a_2 C_2^{{N\over 2}-1}(z)
+a_4 C_4^{{N\over 2}-1}(z)
\ee
with
\bee
a_o={N^2-1\over N(N+2)},
\ee
\bee
a_2={4(N+1)\over N(N-2)(N+4)},
\ee
\bee
a_4={24\over N(N^2-4)(N+4)}.
\ee
Thus we obtain, proceeding as in Section 3
\bea
&&\la |s(i)-\e_N|^4\ra_{\infty,{\rm Dir}}=(\psi_o,|s-\e_N|^4
{\cal T}_{i\infty}\delta_{\e_N})\cr&=&
a_o+a_2C_2^{{N\over 2}-1}(1)\Lambda_2(i,\infty)+
a_4C_4^{{N\over 2}-1}(1)\Lambda_2(i,\infty)
\eea
Using now the asymptotics on $\Lambda_l(i,\infty)$ found in Section3, we 
find that the constant terms, as well as the terms $O(i^{2-D})$, cancel 
and we are left with an expression $O(i^{2(2-D)})$ as claimed. \endproof 

{\it Remark:} By developing the asymptotics of $\Lambda_l(i,\infty)$ 
further, one could presumably push the percolation threshold down to 2. 
Alternatively, percolation for all $D>2$ would follow also from the fact 
that the kernel ${\cal T}_{i\infty}(\e_n,\s)$ decays as a Gaussian with
variance $\la |s(i)-\e_N|^2\ra_{\infty,{\rm Dir}}$.
    %Percolation Properties
\section{Conclusions}
\setcounter{equation}{0}
The main  purpose of this study of a solvable familiy of models was to 
test certain ideas of Patrascioiu and Seiler concerning the 
two-dimensional $O(N)$ spin models. Let us review the outcome.

The central thesis of Patrascioiu and Seiler put forward in 
numerous publications (see for instance \cite{PSperc1, PSperc2, PSperc3}
and references therein) is that there is no fundamental qualitative 
difference between the `abelian' case $N=2$ and the `nonabelian' one 
$N>2$. This is fully confirmed in the solvable models studied here: 
critical behavior depends only on the dimension parameter $D$, with the 
models becoming critical at $D=2$ independent of $N$. 

Another point stressed by Patrascioiu and Seiler \cite{Pprl54} is
the fact that perturbation theory is ambiguous in $D\le 2$ and for $N>2$; 
they suggest that the difference found in the perturbative renormalization 
group between the cases $N=2$ and $N>2$ is an artefact of perturbation 
theory. This is again borne out here, as discussed in Section 4.

Finally concerning percolation properties, we find that the `ring 
formation', proposed in \cite{PSperc1, PSperc2, PSperc3} for the `soft 
phase' in $D=2$ actually takes place in the models studied here.

Due to their simplicity these models are, however, lacking certain 
features that the translation invariant  $O(N)$ models possess: due to 
their essentially one-dimensionl nature there is no difference between 
`ring formation' and formation of `islands' typical for the massive high 
temperature phase. In fact for $D=2$ the high temperature phase has been 
eliminated altogether, in accordance with the fact the models are more ordered 
than the translation invariant ones. So this unfortunately eliminates the 
possibility of studying critical behavior in $\beta$. 

In our view the models studied here give support for the ideas of
Patrascioiu and Seiler, but of course they cannot provide anything like a 
proof of them.

%\input app.tex %Appendix A

  %References
\end{document}